\documentclass[twocolumn,floats,floatfix,superscriptaddress,aps,pra]{revtex4}
\usepackage{eurosym}
\usepackage{amsfonts}
\usepackage{amssymb}
\usepackage{amsmath}
\usepackage{graphicx}
\usepackage{bm}
\usepackage{color}
\usepackage{xcolor}
\usepackage{epsfig}
\usepackage{ifthen}
\usepackage{subfigure}
\usepackage{physics}
\usepackage[hidelinks, colorlinks=true,linkcolor=blue,citecolor=blue,filecolor=blue,urlcolor=blue]{hyperref}

\begin{document}

\title{Complete and robust population transfer between the two ground states
of a three-state loop quantum system by amplitude composite pulse control}
\author{Peter Chernev}
\affiliation{Center for Quantum Technologies, Department of Physics, Sofia University,
James Bourchier 5 blvd., 1164 Sofia, Bulgaria}
\author{Andon A. Rangelov}
\affiliation{Center for Quantum Technologies, Department of Physics, Sofia University,
James Bourchier 5 blvd., 1164 Sofia, Bulgaria}

\begin{abstract}
This work presents a method for achieving complete, robust, and efficient population transfer between the two ground states in a three-level loop quantum system. The approach utilizes composite pulse sequences by effectively mapping the three-state system onto an equivalent two-level system. This transformation allows the use of broadband composite pulses designed initially for conventional two-state quantum systems. Unlike traditional implementations, the composite pulses in the three-level system are not controlled through phase adjustments; instead, they are realized via the amplitude ratio of the Rabi frequencies.
\end{abstract}

\maketitle

\section{Introduction}

Three-level quantum systems play a crucial role in many areas of quantum physics. In these systems, optical transitions governed by standard electric dipole selection rules create straightforward coupling structures between states, typically represented by ladder, $\Lambda$, or $V$ configurations \cite{Allen-Eberly,Shore}. When an extra interaction directly links the two end states of such a three-level chain, a closed-loop configuration emerges, as shown in Fig.~\ref{fig1}. This kind of interaction usually violates electric dipole selection rules, which only permit transitions between states of opposite parity. Nonetheless, it can be achieved through alternative means, such as two-photon optical processes or microwave-driven transitions between hyperfine states. This loop configuration is the most basic discrete quantum system capable of displaying complex probability amplitude interference effects, which continues to attract attention \cite{Carroll,Kosachiov,Unanyan,Fleischhauer,Kral}. A notable example by Carroll and Hioe \cite{Carroll} examined a three-level loop for which they derived analytical solutions for the state amplitudes under the influence of three resonant laser pulses with different temporal profiles. In their scenario, two of the couplings were real-valued, and the third was purely imaginary. This particular setup, benefiting from SU(2) symmetry, allows for the reduction of the three-level loop to an effective two-level system.

In parallel, the composite pulse technique—initially developed for nuclear magnetic resonance (NMR) \cite{Levitt and Freeman,Levitt,Freeman,Tycko1983,Tycko1985,Cho,Wimperis}—offers a highly effective method for precise quantum state control. Rather than applying a single pulse to drive a transition between two states, this method uses a carefully designed sequence of pulses with specific phases, allowing precise shaping of the excitation profile. It enables nearly perfect population inversion and is resilient against variations in critical system parameters, such as pulse strength and detuning. Consequently, this approach merges the accuracy of single $\pi$-pulse excitation with the robustness associated with adiabatic methods.

In this work, we combine the SU(2)-symmetric reduction of the three-level loop system to a two-level effective system \cite{Carroll} with the use of composite pulse sequences to establish a reliable scheme for full population transfer between the two ground states of the loop. The propagator required in this effective two-level system corresponds to a phase gate, which we implement using robust composite phase gates previously developed \cite{Torosov2014}. Unlike standard approaches where control is exerted through phase modulation, the composite pulses in our three-level framework are governed by the amplitude ratio between the Rabi frequencies.

The paper is structured as follows. We begin with an overview of the composite pulse theory in two-level systems and show how a composite phase gate can be constructed. We then present the general Hamiltonian for the three-state loop and identify a specific scenario where it simplifies to an effective two-state system.

Afterward, we revisit the full propagator of the three-level loop, derive the conditions for complete population transfer, and demonstrate how implementing a phase gate in the effective system ensures robust state transfer. Finally, we validate the approach through numerical simulations and conclude with a summary of our results.

\section{Robust composite phase gate in two-state quantum system}

We begin by examining a two-level quantum system governed by the Hamiltonian \cite{Allen-Eberly,Shore}: 
\begin{equation}
\mathbf{H}(t)=\frac{\hbar }{2}%
\begin{bmatrix}
-\Delta (t) & \Omega (t) \\ 
\Omega (t) & \Delta (t)%
\end{bmatrix}%
,  \label{Hamiltonian}
\end{equation}%
where the system evolves according to the time-dependent Schr\"{o}dinger equation: 
\begin{equation}
i\hbar \partial _{t}\mathbf{c}(t)=\mathbf{H}(t)\mathbf{c}(t),
\label{Schrodinger}
\end{equation}%
with the state vector $\mathbf{c}(t)=[c_{1}(t),c_{2}(t)]^{T}$ representing the probability amplitudes of the two quantum states. Here, $\Omega(t)$ denotes the Rabi frequency and $\Delta(t)$ is the detuning. The system’s time evolution is described by the propagator $\mathbf{U}$, which connects the state at initial time $t_i$ to the state at final time $t_f$: 
\begin{equation}
\mathbf{c}(t_{f})=\mathbf{U}(t_{f},t_{i})\mathbf{c}(t_{i}).
\end{equation}%
This general SU(2) propagator can be expressed using the Cayley-Klein complex coefficients $a$ and $b$, which satisfy $|a|^{2}+|b|^{2}=1$ \cite{CP1}: 
\begin{equation}
\mathbf{U}=%
\begin{bmatrix}
a & b \\ 
-b^{\ast } & a^{\ast }%
\end{bmatrix}%
.  \label{Cayley}
\end{equation}%
When a phase shift is applied to the Rabi frequency, $\Omega \rightarrow \Omega \,\text{e}^{i\phi}$, the propagator becomes: 
\begin{equation}
\mathbf{U}_{\phi }=%
\begin{bmatrix}
a & b\,\text{e}^{i\phi } \\ 
-b^{\ast }\,\text{e}^{-i\phi } & a^{\ast }%
\end{bmatrix}%
.  \label{Cayley-Klein}
\end{equation}%
Under resonant driving ($\Delta = 0$), the coefficients simplify to $a=\cos(\mathcal{A}/2)$ and $b=-i\sin(\mathcal{A}/2)$, where the pulse area is defined as $\mathcal{A} = \int_{t_i}^{t_f} \Omega(t)\, \text{d}t$. The transition probability from the initial to the excited state is then given by:
\begin{equation}
P_{1\rightarrow 2}=\sin^2(\mathcal{A}/2).
\end{equation}%
A resonant $\pi$ pulse ($\mathcal{A} = \pi$) yields complete population transfer. Replacing this single $\pi$ pulse with a composite sequence of pulses improves robustness against fluctuations in the pulse area \cite{Levitt and Freeman,Levitt,Freeman}. The overall propagator of such a sequence is the product of the individual propagators:
\begin{equation}
\mathbf{U}^{(N)}=\mathbf{U}_{\phi _{N}}(\mathcal{A}_{N})\mathbf{U}_{\phi
_{N-1}}(\mathcal{A}_{N-1})\cdots \mathbf{U}_{\phi _{2}}(\mathcal{A}_{2})%
\mathbf{U}_{\phi _{1}}(\mathcal{A}_{1}),  \label{propagator}
\end{equation}%
where $\mathcal{A}_j$ and $\phi_j$ represent the pulse area and phase of the $j$-th pulse $(j = 1, 2, ..., N)$. The goal is to achieve complete and robust population transfer between states $|1\rangle$ and $|2\rangle$, while reducing sensitivity to pulse area variations.

To simplify the analysis, we assume all pulses have equal area, $\mathcal{A}_{1} = \mathcal{A}_{2} = \cdots = \mathcal{A}_{N} = \mathcal{A}$, and consider an odd number of pulses, $N = 2n + 1$, although this is not strictly required. Moreover, we impose time-reversal symmetry on the pulse sequence by requiring that the phases satisfy $\phi_{k} = \phi_{N+1-k}$. Since the global phase has no physical effect, we fix $\phi_1 = \phi_N = 0$, leaving $n$ independent phase parameters.

Next, we calculate the composite propagator (\ref{propagator}) and impose conditions such that the first $n$ non-zero derivatives of the matrix element $U_{11}^{(N)}$ with respect to $\mathcal{A}$ vanish at $\mathcal{A} = \pi$. This leads to a system of $n$ nonlinear equations for the $n$ unknown phases. The symmetry condition, $\phi_k = \phi_{N+1-k}$, ensures that either all even or all odd derivatives vanish, allowing the first $2n$ derivatives to be eliminated:
\begin{equation}
\left[ \partial _{\mathcal{A}}^{k}U_{11}^{\left( N\right) }\right] _{%
\mathcal{A}=\pi }=0\ \ \ \left( k=1,2,...,N-2\right) .
\label{nullify the first 2n derivatives}
\end{equation}%
An analytical expression for these optimal phases for any number of pulses was derived by Torosov et al. \cite{CP1}:
\begin{equation}
\phi _{k}^{\left( N\right) }=\left( N+1-2\left\lfloor \frac{k+1}{2}%
\right\rfloor \right) \left\lfloor \frac{k}{2}\right\rfloor \frac{\pi }{2N},
\label{rotation angles}
\end{equation}%
where $k=1,2,...,N$ and $\left\lfloor x \right\rfloor$ denotes the floor function (integer part of $x$). This composite sequence cancels the first $2N - 1$ derivatives of the transition probability with respect to pulse area at $\mathcal{A} = \pi$. The resulting transition probability is:
\begin{equation}
P_{1\rightarrow 2} \approx 1 - \cos^{2N}\left( \mathcal{A}/2 \right),
\end{equation}%
which approaches unity for large $N$, except at values of $\mathcal{A}$ that are even multiples of $\pi$. Table~\ref{Table 1} lists the explicit phase values $\phi_k$ for several composite pulse sequences, as determined by Eq.~(\ref{rotation angles}).
\begin{table}[htb]
\begin{tabular}{cccccccccccccccccc}
\hline\hline
$N$ & \multicolumn{17}{c}{Phases $\phi _{k}$ (in units of $\pi /N$)} \\ 
\hline
3 &  &  &  &  &  &  &  & 0 & 1 & 0 &  &  &  &  &  &  &  \\ 
5 &  &  &  &  &  &  & 0 & 2 & 1 & 2 & 0 &  &  &  &  &  &  \\ 
7 &  &  &  &  &  & 0 & 3 & 2 & 4 & 2 & 3 & 0 &  &  &  &  &  \\ 
9 &  &  &  &  & 0 & 4 & 3 & 6 & 4 & 6 & 3 & 4 & 0 &  &  &  &  \\ 
11 &  &  &  & 0 & 5 & 4 & 8 & 6 & 9 & 6 & 8 & 4 & 5 & 0 &  &  &  \\ 
13 &  &  & 0 & 6 & 5 & 10 & 8 & 12 & 9 & 12 & 8 & 10 & 5 & 6 & 0 &  &  \\ 
15 &  & 0 & 7 & 6 & 12 & 10 & 15 & 12 & 16 & 12 & 15 & 10 & 12 & 6 & 7 & 0 & 
\\ \hline\hline
\end{tabular}%
\caption{Phases of composite pulses for different numbers of $N$ pulses in
the series. }
\label{Table 1}
\end{table}

The composite pulses discussed so far are optimized for resonant conditions ($\Delta = 0$), but the same design principles can be extended to non-resonant scenarios ($\Delta \neq 0$). This extension allows the creation of pulse sequences that are simultaneously insensitive to variations in both the pulse area $\mathcal{A}$ and the detuning $\Delta$. The method for determining the phases remains unchanged; however, the conditions now involve setting derivatives of the propagator $U^{(N)}$ with respect to both parameters to zero:
\begin{equation}
\left[ \frac{\partial ^{k_{1}}\partial ^{k_{2}}}{\partial ^{k_{1}}\mathcal{A}%
\partial ^{k_{2}}\Delta }\right] _{\left( \mathcal{A}=\pi ,\Delta =0\right)
}U_{11}^{\left( N\right) }=0,
\end{equation}%
where the integers $k_{1}$ and $k_{2}$ determine the level of flatness in response to changes in $\mathcal{A}$ and $\Delta$, respectively. These sequences, known as universal composite pulses, were developed by Genov et al. \cite{Genov}. Table~\ref{Table 2} shows the explicitly calculated phase values $\phi_k$ for a few such sequences.

\begin{table}[htb]
\begin{tabular}{cc}
\hline\hline
\ $N$ & Phases $\phi _{k}$ \\ \hline
$3$ & $\left( 0,1,0\right) \pi /2$ \\ 
$5$ & $\left( 0,5,2,5,0\right) \pi /6$ \\ 
$7$ & $\left( 0,11,10,17,10,11,0\right) \pi /12$ \\ 
$13$ & $\left( 0,9,42,11,8,37,2,37,8,11,42,9,0\right) \pi /24$ \\ 
\hline\hline
\end{tabular}%
\caption{Phases of universal composite pulses for different numbers of $N$
pulses in the series. }
\label{Table 2}
\end{table}

Having established how to construct robust composite pulse sequences capable of achieving complete population transfer, we now demonstrate how to implement a phase gate using two such pulses. The phase gate is defined as:
\begin{equation}
\Phi = 
\begin{bmatrix}
e^{i\beta /2} & 0 \\ 
0 & e^{-i\beta /2}%
\end{bmatrix}%
,  \label{phase gate}
\end{equation}
and can be realized by applying two consecutive composite pulses with a specific relative phase. This concept is based on the method proposed by Torosov et al. \cite{Torosov2014}.

As a simple example, consider two single pulses where the second pulse is applied with a relative phase shift $\phi$ compared to the first. By using the single-pulse propagators from Eqs.~(\ref{Cayley}) and (\ref{Cayley-Klein}), the total propagator becomes:
\begin{align}
\mathbf{U}_{\text{tot}} &= \mathbf{U}(\phi )\mathbf{U}(0)  \notag \\
& = 
\begin{bmatrix}
a^{2}-|b|^{2}e^{i\phi } & ab+a^{\ast }be^{i\phi } \\ 
-a^{\ast }b^{\ast }-ab^{\ast }e^{-i\phi } & a^{\ast 2}-|b|^{2}e^{-i\phi }%
\end{bmatrix}%
.  \label{sequence}
\end{align}
This equation illustrates the basic mechanism of the phase gate: if each pulse produces complete population transfer (i.e., $a = 0$ and $|b| = 1$), and we choose $\phi = \pi + \beta/2$ for the second pulse, then the combined effect yields:
\begin{equation}
\mathbf{U}_{\text{tot}} = \mathbf{U}(\pi +\beta /2)\mathbf{U}(0) = \Phi,
\end{equation}
which exactly matches the target phase gate in Eq.~(\ref{phase gate}).

This gate can thus be implemented using a pair of resonant pulses with $a = \cos(A/2)$ and $b = -i \sin(A/2)$, where $A$ is the pulse area. When $A = \pi$, we achieve $a = 0$, and the resulting propagator is equivalent to the desired phase gate.

Although this construction is conceptually simple and elegant, it shares the same sensitivity limitations as single $\pi$ pulses. These drawbacks can be overcome by replacing the single pulses with broadband composite pulses, such as those listed in Tables~\ref{Table 1} and~\ref{Table 2}, significantly improving the gate’s robustness to parameter fluctuations.

\section{Hamiltonian of the three-state loop system}

\begin{figure}[tb]
\includegraphics[width=0.6\columnwidth]{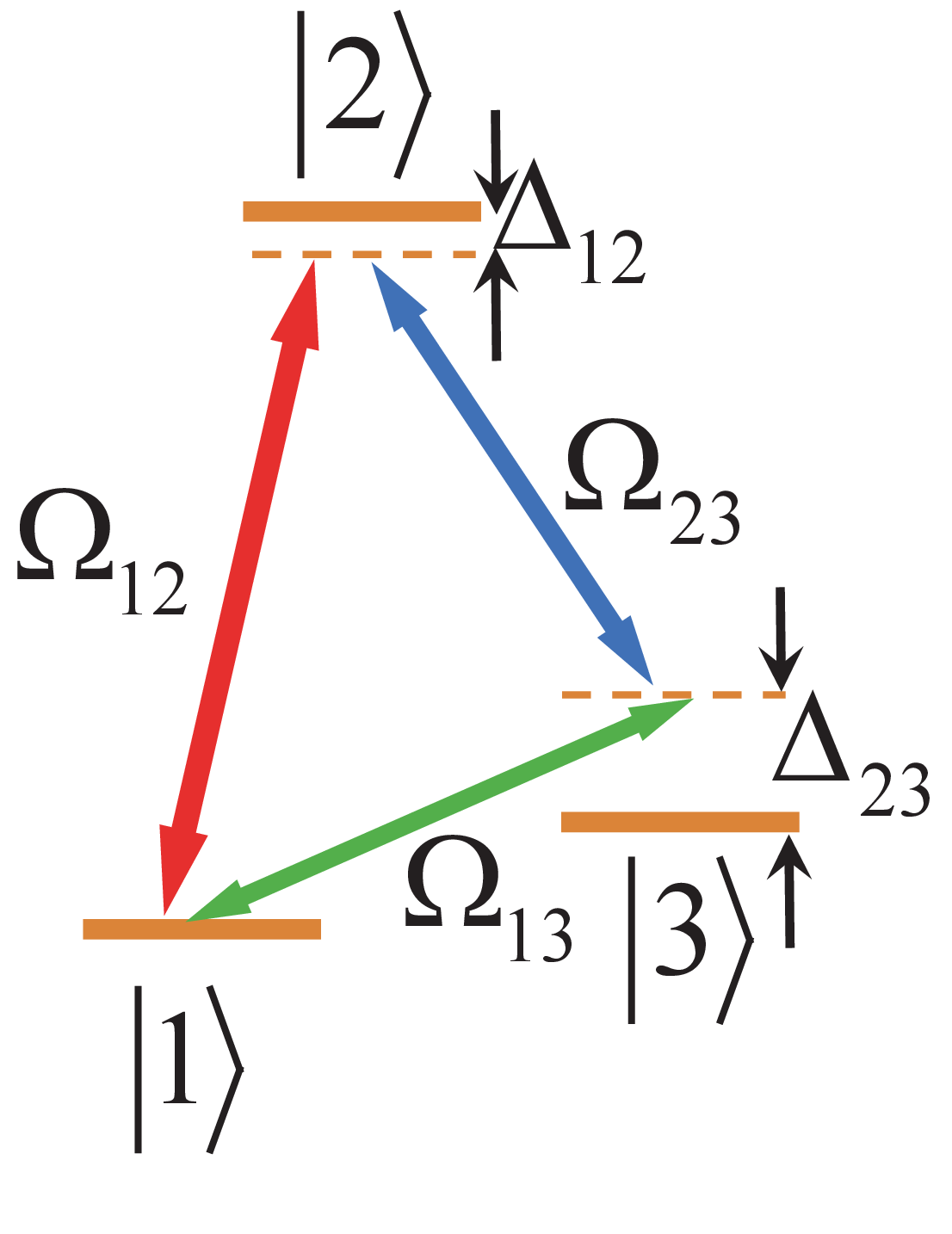}
\caption{(Color online) Coupling scheme of a three-state loop system. The states are connected as follows: $|1\rangle$ and $|2\rangle$ via the Rabi frequency $\Omega_{12}$, $|2\rangle$ and $|3\rangle$ via $\Omega_{23}$, and $|1\rangle$ and $|3\rangle$ via $\Omega_{13}$. Although the levels are arranged such that $|1\rangle$ is the ground state and $|2\rangle$ and $|3\rangle$ are excited, the symmetry of the loop permits the initial population to reside in any of the three states.}
\label{fig1}
\end{figure}

The evolution of a three-state loop system, as depicted in Fig.~\ref{fig1}, is governed by the time-dependent Schr\"{o}dinger equation: 
\begin{equation}
i\hbar \partial _{t}\mathbf{c}(t) = \mathbf{H}(t)\mathbf{c}(t),  \label{Schr}
\end{equation}
where the state vector $\mathbf{c}(t) = [c_{1}(t), c_{2}(t), c_{3}(t)]^{T}$ consists of the probability amplitudes associated with the three quantum states. Under the rotating-wave approximation \cite{Shore}, the most general Hamiltonian in the diabatic basis is given by:
\begin{equation}
\mathbf{H}(t) = \frac{\hbar }{2} 
\begin{bmatrix}
0 & \Omega _{12} & \Omega _{13}^{\ast } \\ 
\Omega _{12}^{\ast } & 2\Delta _{12} & \Omega _{23} \\ 
\Omega _{13} & \Omega _{23}^{\ast } & 2\Delta _{23}%
\end{bmatrix}%
,  \label{H}
\end{equation}
where the complex Rabi frequencies are defined as $\Omega _{12} = |\Omega _{1}|e^{i\varphi _{1}}$, $\Omega _{23} = |\Omega _{2}|e^{i\varphi _{2}}$, and $\Omega _{13} = |\Omega _{3}|e^{i\varphi _{3}}$, representing the couplings between states $|1\rangle \leftrightarrow |2\rangle$, $|2\rangle \leftrightarrow |3\rangle$, and $|1\rangle \leftrightarrow |3\rangle$, respectively. The parameters $\Delta_{12}$ and $\Delta_{23}$ denote the one-photon and two-photon detunings.

Accordingly, the system's dynamics are described by the following Schr\"{o}dinger equation:
\begin{equation}
i\hbar \frac{d}{dt}%
\begin{bmatrix}
c_{1} \\ 
c_{2} \\ 
c_{3}%
\end{bmatrix}%
=\frac{\hbar }{2}%
\begin{bmatrix}
0 & \Omega _{12} & \Omega _{13}^{\ast } \\ 
\Omega _{12}^{\ast } & 2\Delta _{12} & \Omega _{23} \\ 
\Omega _{13} & \Omega _{23}^{\ast } & 2\Delta _{23}%
\end{bmatrix}%
\begin{bmatrix}
c_{1} \\ 
c_{2} \\ 
c_{3}%
\end{bmatrix}%
.  \label{Schroudinger}
\end{equation}

When all elements of the Hamiltonian in Eq.~(\ref{H}) share a common time dependence, it is possible to recast Eq.~(\ref{Schroudinger}) into a form with a time-independent Hamiltonian by applying a suitable time rescaling. This transformation would allow the system's evolution to be described using the eigenvalues and eigenvectors of a static Hamiltonian. However, in this work, we consider a more general situation where at least some of the terms in Eq.~(\ref{H}) vary differently with time. As a result, the Hamiltonian cannot be made time-independent, and the full time-dependent dynamics must be treated explicitly.

\section{Effective two-state system by Carroll and Hioe model}

A special symmetry becomes apparent in the fully resonant case when the phase parameters are set to $\varphi _{1}=0$, $\varphi _{2}=0$, and $\varphi _{3}=\pi /2$ \cite{Carroll}. Physically, this means two of the Rabi frequencies are in phase, while the third is offset by a $\pi/2$ phase. Under these conditions, the dynamics governed by Eq.~(\ref{Schroudinger}) simplify to:
\begin{equation}
i\frac{d}{dt}%
\begin{bmatrix}
c_{1} \\ 
c_{2} \\ 
c_{3}%
\end{bmatrix}%
=\frac{1}{2}%
\begin{bmatrix}
0 & \Omega _{1} & -i\Omega _{3} \\ 
\Omega _{1} & 0 & \Omega _{2} \\ 
i\Omega _{3} & \Omega _{2} & 0%
\end{bmatrix}%
\begin{bmatrix}
c_{1} \\ 
c_{2} \\ 
c_{3}%
\end{bmatrix}%
.  \label{three-state loop Schrodinger}
\end{equation}%
The corresponding Hamiltonian can be written as:
\begin{equation}
\mathbf{H}=\frac{\Omega _{1}}{2}\mathbf{J}_{1}+\frac{\Omega _{2}}{2}\mathbf{J%
}_{2}+\frac{\Omega _{3}}{2}\mathbf{J}_{3},  \label{Lie}
\end{equation}%
where the matrices $\mathbf{J}_{1},\mathbf{J}_{2}$, and $\mathbf{J}_{3}$ are defined by \cite{Carroll,Hioe}:
\begin{eqnarray}
\mathbf{J}_{1} &=&%
\begin{bmatrix}
0 & 1 & 0 \\ 
1 & 0 & 0 \\ 
0 & 0 & 0%
\end{bmatrix}%
, \\
\mathbf{J}_{2} &=&%
\begin{bmatrix}
0 & 0 & 0 \\ 
0 & 0 & 1 \\ 
0 & 1 & 0%
\end{bmatrix}%
, \\
\mathbf{J}_{3} &=&%
\begin{bmatrix}
0 & 0 & -i \\ 
0 & 0 & 0 \\ 
i & 0 & 0%
\end{bmatrix}%
.
\end{eqnarray}%
These matrices satisfy the SU(2) algebra via the commutation relations \cite{Carroll,Hioe}:
\begin{equation}
\left[ \mathbf{J}_{1},\mathbf{J}_{2}\right] =i\mathbf{J}_{3},\quad \left[ 
\mathbf{J}_{2},\mathbf{J}_{3}\right] =i\mathbf{J}_{1},\quad \left[ \mathbf{J}%
_{3},\mathbf{J}_{1}\right] =i\mathbf{J}_{2}.
\end{equation}%
Thus, they form a valid representation of the angular momentum operators for a spin-1 system \cite{Hioe}, and the three-state loop follows the SU(2) symmetry structure \cite{Carroll,Hioe}.

In this symmetric configuration, Carroll and Hioe \cite{Carroll} showed that the three-state system effectively reduces to a two-state system, governed by:
\begin{equation}
i\frac{d}{dt}%
\begin{bmatrix}
c_{1} \\ 
c_{2}%
\end{bmatrix}%
=\frac{1}{2}%
\begin{bmatrix}
\Omega _{3} & \Omega _{1}-i\Omega _{2} \\ 
\Omega _{1}+i\Omega _{2} & -\Omega _{3}%
\end{bmatrix}%
\begin{bmatrix}
c_{1} \\ 
c_{2}%
\end{bmatrix}%
.
\end{equation}%
Here, the complex Rabi frequency is defined as $\Omega = \Omega _{1} + i \Omega _{2}$, and the effective detuning is $\Delta = \Omega _{3}$. If we let $\Omega _{1} = \Omega _{0} \cos \varphi$ and $\Omega _{2} = \Omega _{0} \sin \varphi$, we can express the Rabi frequency as $\Omega = \Omega _{0} e^{i\varphi}$, where $\Omega_0$ is the amplitude and $\varphi$ its phase. This shows that the phase control in the effective two-state model corresponds to a specific ratio of the Rabi frequencies $\Omega_1$ and $\Omega_2$ in the full three-state system.

\begin{figure*}[htb]
\includegraphics[width=140mm,angle=0]{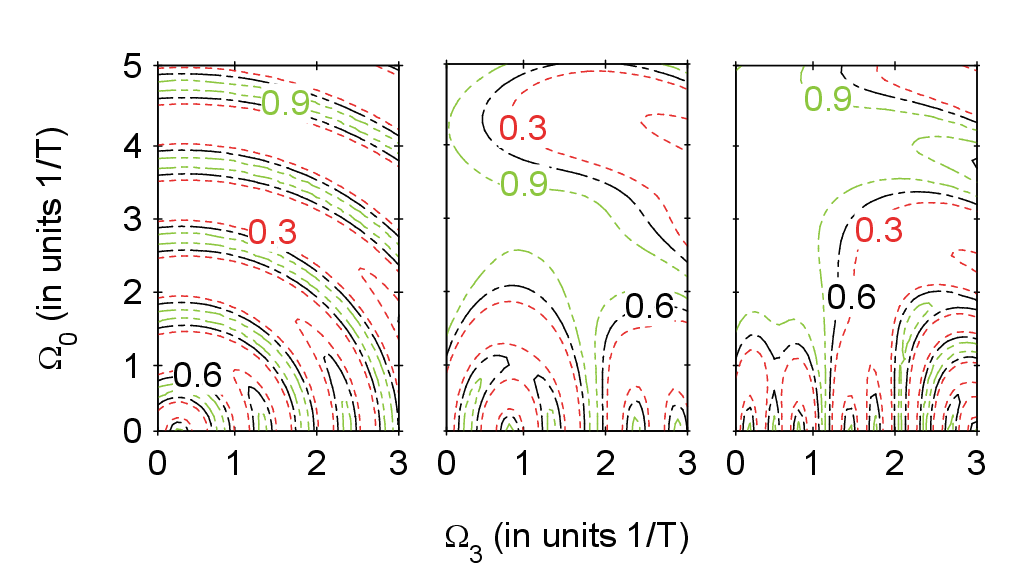}
\caption{(Color online) Contour plot with isoline of the population at state 
$|3\rangle $ $(P_{3}=|c_{3}(\infty )|^{2})$ as a function of the Rabi
frequency amplitudes $\Omega _{0}$ and $\Omega _{3}$. For constant Rabi
frequencies (left frame), six composite pulses (middle frame), and ten
composite pulses (right frame). The pulse shapes are rectangular as given in
Eq.(\protect\ref{rectangular_pulse}). The parameters $\protect\varphi _{k}$
are specified as $(0,\protect\pi /2,0,5\protect\pi /4,7\protect\pi /4,5%
\protect\pi /4)$ for the six-pulse case and $(0,5\protect\pi /6,\protect\pi %
/3,5\protect\pi /6,0,5\protect\pi /4,25\protect\pi /12,19\protect\pi /12,25%
\protect\pi /12,5\protect\pi /4)$ for the ten-pulse case. All curves are
numerically calculated from Eq. (\protect\ref{three-state loop Schrodinger}%
). }
\label{fig3}
\end{figure*}

The propagator of the reduced two-level system can again be described using the Cayley-Klein parameters $a$ and $b$ from Eq.~(\ref{Cayley}). By applying the Carroll-Hioe transformation \cite{Carroll}, the full three-state loop propagator becomes:
\begin{widetext}
\begin{equation}
\mathsf{U}^{\left( 3\right) }=%
\begin{bmatrix}
\frac{1}{2}\left( a^{2}+b^{2}+a^{\ast 2}+b^{\ast 2}\right)  & ab-a^{\ast
}b^{\ast } & \frac{i}{2}\left( b^{2}-a^{2}+a^{\ast 2}-b^{\ast 2}\right)  \\ 
ba^{\ast }-ab^{\ast } & \left\vert a\right\vert ^{2}-\left\vert b\right\vert
^{2} & i\left( ba^{\ast }+ab^{\ast }\right)  \\ 
\frac{i}{2}\left( a^{2}+b^{2}-a^{\ast 2}-b^{\ast 2}\right)  & i\left(
ab+a^{\ast }b^{\ast }\right)  & \frac{1}{2}\left( a^{2}-b^{2}+a^{\ast
2}-b^{\ast 2}\right) 
\end{bmatrix}%
.  \label{U3 CH}
\end{equation}
\end{widetext}
Assuming the system starts in state $\left|1\right\rangle$, i.e., $c_1(t_i)=1$ and $c_2(t_i)=c_3(t_i)=0$, the final state populations are:
\begin{eqnarray}
P_{1} &=&\left\vert c_{1}(t)\right\vert ^{2}=\frac{\left\vert
a^{2}+b^{2}+a^{\ast 2}+b^{\ast 2}\right\vert ^{2}}{4}, \\
P_{2} &=&\left\vert c_{2}(t)\right\vert ^{2}=\left\vert ba^{\ast }-ab^{\ast
}\right\vert ^{2}, \\
P_{3} &=&\left\vert c_{3}(t)\right\vert ^{2}=\frac{\left\vert
a^{2}+b^{2}-a^{\ast 2}-b^{\ast 2}\right\vert ^{2}}{4}.
\end{eqnarray}%
In the case where $a = e^{i\beta/2}$ and $b = 0$, we get:
\begin{eqnarray}
P_{1} &=&\frac{\left\vert e^{i\beta }+e^{-i\beta }\right\vert ^{2}}{4}=\cos
^{2}\beta , \\
P_{2} &=&0, \\
P_{3} &=&\frac{\left\vert e^{i\beta }-e^{-i\beta }\right\vert ^{2}}{4}=\sin
^{2}\beta .
\end{eqnarray}%
This means that constructing a propagator of the form:
\begin{equation}
\mathbf{U}=\left[ 
\begin{array}{cc}
e^{i\beta /2} & 0 \\ 
0 & e^{-i\beta /2}%
\end{array}%
\right] ,  \label{U}
\end{equation}%
in the effective two-state system results in a coherent superposition between states $\left|1\right\rangle$ and $\left|3\right\rangle$ in the original three-level system. This matrix corresponds to the phase gate defined in Eq.~(\ref{phase gate}). Setting $\beta = \pi/2$ yields complete population transfer. Since broadband composite phase gates have been developed previously \cite{Torosov2014} and summarized in the prior section, these can now be directly applied within the effective two-state framework to achieve robust and complete population transfer in the three-state loop system.

\section{Numerical simulations}

To begin simulations, we emphasize that since this method operates under exact resonance conditions, the specific form of the pulse shape is not critical. While our examples utilize rectangular pulse profiles, defined as:
\begin{widetext}
\begin{subequations}
\label{rectangular_pulse}
\begin{eqnarray}
\Omega _{1}(t) &=&\Omega _{0}\sum_{k=0}^{N}\cos \varphi _{k}\left( UnitStep(t-kT)-{UnitStep}(t-(k+1)T)\right), \\
\Omega _{2}(t) &=&\Omega _{0}\sum_{k=0}^{N}\sin \varphi _{k}\left( UnitStep(t-kT)-UnitStep(t-(k+1)T)\right), \\
\Omega _{3}(t) &=&\Omega _{3},
\end{eqnarray}
\end{subequations}
\end{widetext}
where $k$ is an integer, $T$ denotes the duration of each pulse, $N$ is the total number of pulses, and $\Omega_0$ and $\Omega_3$ are constant amplitudes. Nevertheless, other pulse shapes—such as Gaussian or smoothly varying ones—can be employed with comparable results.

Figure~\ref{fig3} shows contour plots of the final population in state $|3\rangle$, given by $P_{3} = |c_{3}(\infty)|^{2}$, for systems driven by six and ten composite pulses, alongside the case of unmodulated, constant Rabi frequencies. These results are obtained by numerically integrating Eq.~(\ref{three-state loop Schrodinger}). The plots reveal that increasing the number of composite pulses significantly enhances the robustness and effectiveness of population transfer, yielding a broader region of high-fidelity transition. This highlights the superior performance of composite pulses compared to constant fields.

\begin{figure*}[htb]
\includegraphics[width=120mm,angle=0]{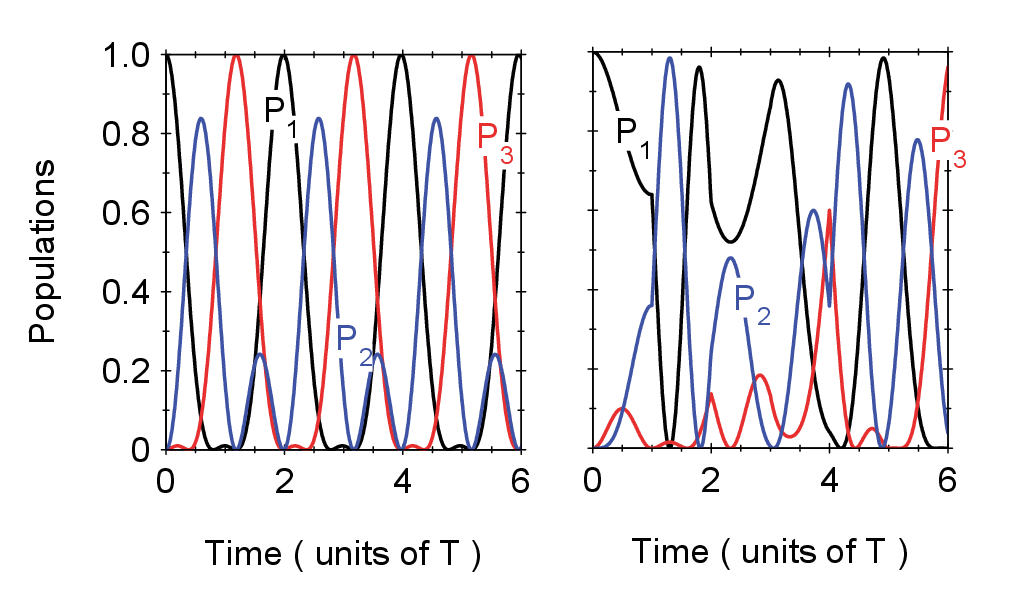}
\caption{(Color online) Time evolution of state populations for constant Rabi frequencies (left) and a six-pulse composite sequence (right). Pulses follow the rectangular shape defined in Eq.~(\protect\ref{rectangular_pulse}) with $\Omega_0 = 3/T$ and $\Omega_3 = 1/T$. The phase values $\varphi_k$ for the six-pulse sequence are $(0, \protect\pi/2, 0, 5\protect\pi/4, 7\protect\pi/4, 5\protect\pi/4)$. All results are obtained by numerically solving Eq.~(\protect\ref{three-state loop Schrodinger}).}
\label{fig2}
\end{figure*}

Figure~\ref{fig2} presents the time evolution of population probabilities for two different cases: constant Rabi frequencies (left panel) and a six-pulse composite sequence (right panel). The parameters are selected to ensure effective population transfer with the six-pulse sequence, based on the parameter space $(\Omega_0, \Omega_3)$ explored in Fig.~\ref{fig3}.

\section{Conclusion}

In this work, we have introduced a method for realizing complete population transfer between the two ground states of a three-state loop quantum system. The technique leverages a specific configuration of the loop system, with its Hamiltonian described by Eq.~(\ref{Lie}), and takes advantage of the fact that this system can be mapped onto an effective two-level model governed by the Hamiltonian in Eq.~(\ref{Hamiltonian}).

Within this reduced two-state framework, achieving full population transfer in the original three-level system is equivalent to implementing a phase gate with a phase of $\pi/2$. To enhance the robustness of the transition against fluctuations in interaction parameters, we replace single pulses with composite pulse sequences. These sequences introduce phase control in the effective two-state system, which corresponds, in the original three-level configuration, to a specific ratio between Rabi frequencies. 

A key feature of our approach is that, unlike conventional composite pulse schemes that directly manipulate phases, control in the three-level loop system is exerted through the amplitude ratios of the Rabi frequencies. Furthermore, we note that the well-known three-state $\Lambda$ system represents a special case of the loop configuration, realized when the coupling $\Omega_3$ is set to zero.

Beyond enabling full population transfer, our method also allows the preparation of arbitrary superpositions of the ground states. These superpositions can be generated by designing custom phase gates using composite pulses within the effective two-level model, expanding the range of controllable operations in such quantum systems.
\acknowledgements

This research is partially supported by the Bulgarian national plan for
recovery and resilience, contract BG-RRP-2.004-0008-C01 SUMMIT: Sofia
University Marking Momentum for Innovation and Technological Transfer,
project number 3.1.4.

\end{document}